\RequirePackage{fix-cm}
\documentclass[graybox]{svmult}
\usepackage{mathptmx}       % selects Times Roman as basic font
\usepackage{helvet}         % selects Helvetica as sans-serif font
\usepackage{courier}        % selects Courier as typewriter font
\usepackage[bottom]{footmisc}% places footnotes at page bottom
\usepackage{amsmath}
\usepackage{amssymb}
\usepackage{mathbbol}
\newcommand{\HC}{\mathcal{H}}
\newcommand{\LC}{\mathcal{L}}
\newcommand{\dd}{\mathrm{d}}
\newcommand{\onehalf}{{\textstyle\frac{1}{2}}}
\newcommand{\quarter}{{\textstyle\frac{1}{4}}}
\newcommand{\pfrac}[2]{\frac{\partial{#1}}{\partial{#2}}}
\newcommand{\ppfrac}[3]{\frac{\partial^{2}{#1}}{\partial{#2}\partial{#3}}}
\newcommand{\pffrac}[3]{\pfrac{#1}{\left(\pfrac{#2}{#3}\right)}}

\newcommand{\rmi}{\mathrm{i}}
\newcommand{\CO}{\mathcal{O}}
\newcommand{\btau}{\pmb{\tau}}
\newcommand{\balpha}{\pmb{\alpha}}
\let\overline=\bar
\begin{document}
\bibliographystyle{spphys}
\title*{Covariant Hamiltonian representation of Noether's theorem
and its application to\\ U$(N)$ and SU$(N)$ gauge theories}
\titlerunning{Hamiltonian representation of Noether's theorem}
% Use for an abbreviated version of
\author{J\"{u}rgen Struckmeier, Horst St\"{o}cker, and David Vasak}
\authorrunning{J.~Struckmeier, H.~St\"{o}cker, D.~Vasak}
% for an abbreviated version of
\institute{J.~Struckmeier \at
Frankfurt Institute for Advanced Studies (FIAS), Ruth-Moufang-Strasse~1, 60438~Frankfurt\\
Goethe Universit\"{a}t, Max-von-Laue-Strasse~1, 60438~Frankfurt am Main, Germany\\
GSI Helmholtzzentrum f\"{u}r Schwerionenforschung GmbH, Planckstr.~1, 64289~Darmstadt\\
\email{struckmeier@fias.uni-frankfurt.de}
\and H.~St\"{o}cker \at
Judah Eisenberg Professor Laureatus\\
GSI Helmholtzzentrum f\"{u}r Schwerionenforschung GmbH, Planckstr.~1, 64289~Darmstadt\\
Frankfurt Institute for Advanced Studies (FIAS), Ruth-Moufang-Strasse~1, 60438~Frankfurt\\
Institute of Theoretical Physics (ITP), Goethe University, Max-von-Laue-Strasse~1,\\ 60438~Frankfurt am Main, Germany\\
\email{stoecker@fias.uni-frankfurt.de}
\and D.~Vasak \at
Frankfurt Institute for Advanced Studies (FIAS), Ruth-Moufang-Strasse~1, 60438~Frankfurt\\
\email{vasak@fias.uni-frankfurt.de}}
\maketitle
\abstract{
We present the covariant Hamiltonian representation of Noether's theorem
and its application to Yang-Mills gauge theory.
As the starting point, we re-formulate our previous presentation of the
canonical Hamiltonian derivation of Noether's theorem~\cite{StrRei12},
distinguishing admissible canonical transformations from symmetries of a
specified Hamiltonian.
The Noether currents of U$(1)$ and U$(N)$ gauge theories, and their
restriction to SU$(N)$, are derived from the respective infinitesimal
generating functions of the pertinent symmetry transformations.
The resulting conservation relations are distinguished from the complete
field equations, which require the chosen kinetic Hamiltonian and a
variational treatment of the antisymmetric gauge momenta.}
%\pacs{11.10.Ef, 11.15.Kc}
\section{Introduction}
Noether's theorem establishes in the realm of the Hamilton-Lagrange
description of continuum dynamics the correlation
of a conserved current with a particular symmetry transformation
that preserves the form of the Hamiltonian of the given system.
Although usually derived in the Lagrangian formalism~\cite{noether18,saletan98},
Noether's theorem also has a natural formulation in covariant Hamiltonian field theory.
Canonical transformations preserve the canonical form of the
action principle, with the Hamiltonian transformed accordingly~\cite{struckmeier08}.
A symmetry of a specified Hamiltonian requires, in addition, that the
transformed Hamiltonian have the same functional form as the original one.
For a conserved current that satisfies the admissibility conditions of the
generating-function ansatz used below, the associated infinitesimal
transformation is given by the canonical transformation rules.
Conversely, an admissible canonical symmetry yields a current that is
conserved on solutions of the canonical field equations.
The converse from conservation to symmetry is understood to hold for all
local solutions, with the local-solvability qualification stated in
Sect.~\ref{sec:Noether-fields}.
This gives a covariant Hamiltonian representation of Noether's theorem
within the stated class of fixed-coordinate transformations.
For local gauge symmetries, the corresponding Noether identities must
further be distinguished from the dynamical equations supplied by the
chosen Hamiltonian.
\section{\label{sec:Lag-fields}Lagrangian description of the dynamics of fields}
The realm of classical continuum physics deals
with the dynamics of a system of $N\geq1$ fields $\phi_{I}(x)$
which are functions of space $(x^{1},x^{2},x^{3})$ and
time $t\equiv x^{0}/c$ as the independent variables, $x\equiv(x^{0},x^{1},x^{2},x^{3})$
(see, e.g.\ Greiner, Class.~Electrodyn.~\cite{greiner98}).
Depending on the context of our description, an indexed quantity
may denote as well the complete collection of the respective quantities.
In the first-order Lagrangian description, the state of the system is completely
described by the actual fields $\phi_{I}(x)$ and their $4N$ partial derivatives
$\partial_{\mu}\phi_{I}(x),\,\mu=0,\ldots,3;\,I=1,\ldots,N$.
We assume the dynamical system to be described by a first-order Lagrangian
density $\LC$ which may explicitly depend on the independent variables,
\begin{equation}\label{L-conv-def}
\LC\left(\phi_{I},\partial\phi_{I},x\right).
\end{equation}
Herein, $\partial\phi_{I}$ denotes the complete set of
partial derivatives of $\phi_{I}(x)$.
The Lagrangian density $\LC$ is a local function of the field values and
their first derivatives, whereas the action $S[\phi]$ is a functional of the fields.
We use the Minkowski metric $\eta_{\mu\nu}=\operatorname{diag}(1,-1,-1,-1)$ to raise
and lower space-time indices and use natural units in the following equations.

The space-time evolution of a dynamical system follows from
the \emph{principle of stationary action}: the variation $\delta S$ of the action functional,
\begin{equation}\label{action-int}
S=\int_{R}\LC\left(\phi_{I},\partial\phi_{I},x\right)\dd^{4}x,
\qquad\delta S\stackrel{!}{=}0,
\end{equation}
vanishes for solutions under arbitrary smooth field variations that
vanish on the boundary $\partial R$.
From the calculus of variations~\cite{saletan98}, one finds that
$\delta S=0$ holds exactly if the fields $\phi_{I}$ and their
partial derivatives satisfy the Euler-Lagrange field equations
\begin{equation}\label{elgl}
\pfrac{}{x^{\alpha}}\pfrac{\LC}{(\partial_{\alpha}\phi_{I})}-\pfrac{\LC}{\phi_{I}}=0.
\end{equation}
\section{\label{sec:Ham-fields}Covariant Hamiltonian description
of the dynamics of fields in the DeDonder-Weyl formalism}
In order to derive the equivalent \emph{covariant} Hamiltonian
description of continuum dynamics, we follow the classic approach
of T.~De~Donder and H.~Weyl~\cite{dedonder30,weyl35} in tensor language:
define for each field $\phi_{I}(x)$ a conjugate momentum $4$-vector field $\pi_{I}^{\alpha}(x)$.
Their components are given by
\begin{equation}\label{p-def}
\pi_{I}^{\alpha}=\pfrac{\LC}{(\partial_{\alpha}\phi_{I})}\equiv
\pffrac{\LC}{\phi_{I}}{x^{\alpha}}.
%\pfrac{\LC}{\left(\pfrac{\phi_{I}}{x^{\alpha}}\right)}.
\end{equation}
For each scalar field $\phi_{I}$, the $4$-vectors $\pi_{I}^{\alpha}$ are thus induced by the
Lagrangian $\LC$ as the \emph{dual counterparts} of
the $4$-covectors ($1$-forms) $\partial_{\alpha}\phi_{I}$.
For the entire set of $N$ scalar fields $\phi_{I}(x)$,
this establishes a set of $N$ conjugate $4$-vector fields.
With this definition of the $4$-vectors of canonical momenta
$\pi_{I}(x)$, we now define the Hamiltonian
density $\HC(\phi_{I},\pi_{I},x)$ as the
covariant Legendre transform of the Lagrangian density
$\LC(\phi_{I},\partial\phi_{I},x)$ via
\begin{equation}\label{H-def}
\HC(\phi_{I},\pi_{I},x)=\pi_{J}^{\alpha}\,
\pfrac{\phi_{J}}{x^{\alpha}}-\LC(\phi_{I},\partial\phi_{I},x),
\end{equation}
where summation over the pairs of upper and lower indices is understood.
At this point suppose that $\LC$ is \emph{regular},
hence that the complete $4N\times4N$ Hessian matrix
\begin{displaymath}
W_{I\mu,J\nu}=\ppfrac{\LC}{(\partial_{\mu}\phi_{I})}{(\partial_{\nu}\phi_{J})}
\end{displaymath}
is non-singular.
Non-singularity of its separate diagonal field blocks alone is not sufficient.
This ensures local invertibility of the Legendre map and the local
equivalence of the Lagrangian and Hamiltonian descriptions.
The gauge-field sectors considered below are singular with respect to the full set of potential derivatives.
They will instead be treated by a first-order action with antisymmetric gauge momenta; cf.~\cite{vignolo05}.
The definition of $\HC$ by Eq.~(\ref{H-def}) is referred to in literature as the
``De~Donder-Weyl'' Hamiltonian density~\cite{dedonder30,weyl35}.

Obviously, the dependencies of $\HC$ and $\LC$ on the
$\phi_{I}$ and the $x^{\mu}$ only differ by a sign,
\begin{displaymath}
\pfrac{\HC}{\phi_{I}}=-\pfrac{\LC}{\phi_{I}},\qquad
\left.\pfrac{\HC}{x^{\mu}}\right\vert_{\mathrm{expl}}=
-\left.\pfrac{\LC}{x^{\mu}}\right\vert_{\mathrm{expl}}.
\end{displaymath}
In order to derive the canonical field equations, we calculate
from Eq.~(\ref{H-def}) the partial derivative of $\HC$ with
respect to $\pi_{I}^{\mu}$,
\begin{displaymath}
\pfrac{\HC}{\pi_{I}^{\mu}}=\delta_{\mu}^{\alpha}\delta_{IJ}
\,\pfrac{\phi_{J}}{x^{\alpha}}=\pfrac{\phi_{I}}{x^{\mu}}.
\end{displaymath}
In conjunction with the Euler-Lagrange equation~(\ref{elgl}),
we obtain the set of covariant canonical field equations,
\begin{equation}\label{eq:fgln}
\pfrac{\HC}{\pi_{I}^{\alpha}}=\pfrac{\phi_{I}}{x^{\alpha}},\qquad
\pfrac{\HC}{\phi_{I}}=-\pfrac{\pi_{I}^{\alpha}}{x^{\alpha}}.
\end{equation}
These pairs of first-order partial differential equations
are equivalent to the set of second-order differential
equations of Eq.~(\ref{elgl}).
Provided the Lagrangian density $\LC$ is a Lorentz scalar, the
covariant Legendre transformation~(\ref{H-def}) passes this property to
$\HC$, and the canonical field equations are Lorentz-covariant.
Lorentz symmetry of a specified system additionally requires any prescribed
background quantities to share that symmetry.

From the right hand side of the second canonical
field equation~(\ref{eq:fgln}), we observe that this equation alone
fixes only the \emph{divergence} of the conjugate vector field $\pi_{I}$.
A prescribed divergence-free shift,
\begin{equation}\label{eichung}
\pi_{I}\mapsto\Pi_{I}=\pi_{I}+\eta_{I}(x),\qquad
\pfrac{\eta_{I}^{\alpha}}{x^{\alpha}}=0,
\end{equation}
leaves this divergence unchanged.
It does not, however, generally preserve the full set of canonical equations for a fixed Hamiltonian,
including the first equation of~(\ref{eq:fgln}).
Thus Eq.~(\ref{eichung}) is not by itself a gauge freedom of the given system.
It may instead describe a canonical change of variables if the Hamiltonian is transformed consistently.
\section{\label{sec:Cantra-fields}Canonical transformations in the realm of field dynamics}
Similar to the canonical formalism of point mechanics,
we call a transformation of the fields
$(\phi_{I},\pi_{I})\mapsto(\Phi_{I},\Pi_{I})$
\emph{canonical} if the \emph{form} of the variational principle which
is based on the action functional~(\ref{action-int}) is maintained,
\begin{equation}\label{varprinzip}
\delta\int_{R}\left(\pi_{J}^{\alpha}\,\pfrac{\phi_{J}}{x^{\alpha}}
-\HC(\phi_{I},\pi_{I},x)\right)\dd^{4}x=
\delta\int_{R}\left(\Pi_{J}^{\alpha}\,\pfrac{\Phi_{J}}{x^{\alpha}}
-\HC^{\prime}(\Phi_{I},\Pi_{I},x)\right)\dd^{4}x.
\end{equation}
We implement the requirement~(\ref{varprinzip}) locally by allowing
the \emph{integrands} to differ by the divergence
of a $4$-vector field $F_{1}^{\mu},\mu=0,\ldots,3$
whose variation vanishes on the boundary $\partial R$
of the integration region $R$ within space-time
\begin{displaymath}
\delta\int_{R}\pfrac{F_{1}^{\alpha}}{x^{\alpha}}\dd^{4}x=
\delta\oint_{\partial R}F_{1}^{\alpha}dS_{\alpha}\stackrel{!}{=}0.
\end{displaymath}
The obvious consequence of the form invariance of the
variational principle is the form invariance of the
covariant canonical field equations~(\ref{eq:fgln}).
For the integrands of Eq.~(\ref{varprinzip}), which are
actually the Lagrangian densities $\LC$ and $\LC^{\prime}$,
we thus obtain the condition
\begin{eqnarray}
\LC&=&\LC^{\prime}+\pfrac{F_{1}^{\alpha}}{x^{\alpha}}\label{intbed}\\
\pi_{J}^{\alpha}\pfrac{\phi_{J}}{x^{\alpha}}-
\HC(\phi_{I},\pi_{I},x)&=&\Pi_{J}^{\alpha}
\pfrac{\Phi_{J}}{x^{\alpha}}-\HC^{\prime}(\Phi_{I},\Pi_{I},x)+
\pfrac{F_{1}^{\alpha}}{x^{\alpha}}.\nonumber
\end{eqnarray}
With the definition
$F^{\mu}_{1}\equiv F^{\mu}_{1}(\phi_{I},\Phi_{I},x)$,
we restrict ourselves to a function of exactly
those arguments which now enter into transformation rules
for the transition from the original to the new fields.
The divergence of $F^{\mu}_{1}$ reads, explicitly,
\begin{equation}\label{divF}
\pfrac{F_{1}^{\alpha}}{x^{\alpha}}=
\pfrac{F_{1}^{\alpha}}{\phi_{J}}\pfrac{\phi_{J}}{x^{\alpha}}+
\pfrac{F_{1}^{\alpha}}{\Phi_{J}}\pfrac{\Phi_{J}}{x^{\alpha}}+
{\left.\pfrac{F_{1}^{\alpha}}{x^{\alpha}}\right\vert}_{\mathrm{expl}}.
\end{equation}
The rightmost term denotes the sum over the \emph{explicit}
dependencies of the generating function $F^{\mu}_{1}$ on the $x^{\mu}$.
Comparing the coefficients of Eqs.~(\ref{intbed}) and (\ref{divF}),
we find the local coordinate representation of the field
transformation rules which are induced by the generating function $F^{\mu}_{1}$
\begin{equation}\label{genF1}
\pi_{I}^{\mu}=\pfrac{F_{1}^{\mu}}{\phi_{I}},\qquad
\Pi_{I}^{\mu}=-\pfrac{F_{1}^{\mu}}{\Phi_{I}},\qquad
\HC^{\prime}=\HC+{\left.\pfrac{F_{1}^{\alpha}}
{x^{\alpha}}\right\vert}_{\mathrm{expl}}.
\end{equation}
In contrast to the transformation rule for the Lagrangian
density $\LC$ of Eq.~(\ref{intbed}), the rule for the
Hamiltonian density is determined by the \emph{explicit}
dependence of the generating function $F^{\mu}_{1}$ on the $x^{\mu}$.
Hence, if a generating function does not explicitly
depend on the independent variables, $x^{\mu}$, then the
\emph{value} of the Hamiltonian density is not changed
under the particular canonical transformation emerging thereof.

A type-$2$ generating function instead depends on the
original fields $\phi_{I}$ and the new conjugate fields $\Pi_{I}^{\mu}$.
Where elimination of the appropriate variables is possible, it is related
to a type-1 generating function by the formal Legendre transformation
\begin{equation}\label{legendre1}
F_{2}^{\mu}(\phi_{I},\Pi_{I},x)=
F_{1}^{\mu}(\phi_{I},\Phi_{I},x)+\Phi_{J}\Pi_{J}^{\mu}.
\end{equation}
We thus encounter the set of transformation rules
\begin{equation}\label{genF2}
\pi_{I}^{\mu}=\pfrac{F_{2}^{\mu}}{\phi_{I}},\qquad
\Phi_{I}\delta^{\mu}_{\nu}=\pfrac{F_{2}^{\mu}}{\Pi_{I}^{\nu}},\qquad
\HC^{\prime}=\HC+{\left.\pfrac{F_{2}^{\alpha}}
{x^{\alpha}}\right\vert}_{\mathrm{expl}},
\end{equation}
which can also be derived directly by using
$F_{1}^{\mu}=F_{2}^{\mu}-\Phi_{J}\Pi_{J}^{\mu}$ in the divergence identity~(\ref{intbed}).
This direct derivation does not require a \mbox{type-$1$}
representation on an open domain of independent polymomenta.
In either representation the generating relations must hold simultaneously
and define a locally invertible map on the relevant phase space.
Non-singularity of separate mixed Hessians for fixed $\mu$ is not a sufficient condition.
In particular, the second rule in~(\ref{genF2})
restricts the admissible dependence of $F_{2}^{\mu}$ on the momenta.
We make this restriction explicit for the infinitesimal transformations below.
\section{\label{sec:Noether-fields}Noether's theorem in the Hamiltonian description of field dynamics}
Canonical transformations are defined as the particular subset of general transformations of the fields
$\phi_{I}$ and their conjugate momentum vector fields $\pi_{I}$
which preserve the form of the action functional~(\ref{varprinzip}).
Such a transformation represents a symmetry of the given Hamiltonian
only if its functional form is also preserved.
The associated four-current then has vanishing space-time divergence on solutions.
In the following, we work out the correlation of this
conserved current by means of an \emph{infinitesimal} canonical
transformation of the field variables.
The generating function $F_{2}^{\mu}$ of an \emph{infinitesimal}
transformation differs from that of an \emph{identical}
transformation by an infinitesimal parameter $\epsilon\neq0$ times
an admissible characteristic function $j^{\mu}(\phi_{I},\Pi_{I},x)$:
\begin{equation}\label{gen-infini}
F_{2}^{\mu}(\phi_{I},\Pi_{I},x)=\phi_{J}\,\Pi_{J}^{\mu}+
\epsilon\,j^{\mu}(\phi_{I},\Pi_{I},x).
\end{equation}
Here \emph{admissible} means that the second rule in~(\ref{genF2})
defines the same field variation for every space-time component.
Explicitly,
\begin{displaymath}
\pfrac{j^{\mu}}{\Pi_{I}^{\nu}}=0\quad(\mu\ne\nu),\qquad
\pfrac{j^{\mu}}{\Pi_{I}^{\mu}}=
\pfrac{j^{\nu}}{\Pi_{I}^{\nu}}\quad\hbox{for all }\mu,\nu,
\end{displaymath}
with no summation over the displayed diagonal indices.
For independent polymomentum components on an open domain, these conditions
imply that $j^{\mu}$ is affine in the momenta: its coefficient of
$\Pi_{I}^{\mu}$ is independent of all momenta and is the same for every $\mu$.
Each component may also contain a momentum-independent term.
The resulting near-identity map is locally invertible for sufficiently small $\epsilon$.
Currents outside this admissible class are not covered
by the present fixed-coordinate, derivative-free ansatz.
To first order, $\Pi_{I}^{\mu}$ may be replaced by $\pi_{I}^{\mu}$ within
$j^{\mu}$ and its derivatives, since $\Pi_{I}^{\mu}=\pi_{I}^{\mu}+\CO(\epsilon)$.
The subsequent transformation
rules follow to first order in $\epsilon$ from the general rules~(\ref{genF2}) as
\begin{displaymath}
\Pi_{I}^{\mu}=\pi_{I}^{\mu}-\epsilon\,\pfrac{j^{\mu}}{\phi_{I}},\qquad
\Phi_{I}\,\delta^{\mu}_{\nu}=\phi_{I}\,\delta^{\mu}_{\nu}+
\epsilon\,\pfrac{j^{\mu}}{\pi_{I}^{\nu}},\qquad
\HC^{\prime}=\HC+\epsilon{\left.\pfrac{j^{\alpha}}
{x^{\alpha}}\right\vert}_{\mathrm{expl}},
\end{displaymath}
hence
\begin{equation}\label{gl1}
\delta\pi_{I}^{\mu}=-\epsilon\,\pfrac{j^{\mu}}{\phi_{I}},\qquad
\delta\phi_{I}\,\delta^{\mu}_{\nu}=\epsilon\,\pfrac{j^{\mu}}{\pi_{I}^{\nu}},\qquad
{\delta\HC|}_{\mathrm{CT}}=\epsilon{\left.\pfrac{j^{\alpha}}{x^{\alpha}}\right\vert}_{\mathrm{expl}}.
\end{equation}
As the transformation does not change the independent variables,
$x^{\mu}$, both the original as well as the transformed fields refer
to the same space-time event $x^{\mu}$, hence \mbox{$\delta x^{\mu}=0$}.
With the transformation rules~(\ref{gl1}), the divergence of the
four-vector of characteristic functions $j^{\mu}$ is given by
\begin{eqnarray*}
\epsilon\,\pfrac{j^{\alpha}}{x^{\alpha}}&=&
\epsilon\,\pfrac{j^{\alpha}}{\phi_{I}}\pfrac{\phi_{I}}{x^{\alpha}}+
\epsilon\,\pfrac{j^{\alpha}}{\pi_{I}^{\beta}}\pfrac{\pi_{I}^{\beta}}{x^{\alpha}}+
\epsilon\left.\pfrac{j^{\alpha}}{x^{\alpha}}\right\vert_{\mathrm{expl}}\\
&=&-\delta\pi_{I}^{\alpha}\pfrac{\phi_{I}}{x^{\alpha}}+
\delta\phi_{I}\,\pfrac{\pi_{I}^{\alpha}}{x^{\alpha}}+{\delta\HC|}_{\mathrm{CT}}.
\end{eqnarray*}
The canonical field equations~(\ref{eq:fgln}) apply along the system's space-time evolution.
The derivatives of the fields with respect to the independent
variables may be then replaced accordingly to yield
\begin{displaymath}
\epsilon\,\pfrac{j^{\alpha}}{x^{\alpha}}=-\pfrac{\HC}{\pi_{I}^{\alpha}}\,\delta\pi_{I}^{\alpha}-
\pfrac{\HC}{\phi_{I}}\,\delta\phi_{I}+{\delta\HC|}_{\mathrm{CT}}.
\end{displaymath}
On the other hand, the variation $\delta\HC|_{\mathrm{fields}}$ of the Hamiltonian
due to the variations $\delta\phi_{I}$ and $\delta\pi_{I}$ of the canonical fields is given by
\begin{equation}\label{gl2}
\delta\HC\big|_{\mathrm{fields}}=\pfrac{\HC}{\phi_{I}}\,\delta\phi_{I}+
\pfrac{\HC}{\pi_{I}^{\alpha}}\,\delta\pi_{I}^{\alpha}.
\end{equation}
If the infinitesimal transformation rule $\delta\HC|_{\mathrm{CT}}$
from Eqs.~(\ref{gl1}) coincides identically on the phase-space domain
with $\delta\HC|_{\mathrm{fields}}$ from Eq.~(\ref{gl2}), then the
canonical transformation defines a \emph{symmetry} of the given Hamiltonian.
Equivalently, $\HC^{\prime}(\Phi,\Pi,x)=\HC(\Phi,\Pi,x)$ to first order.
Along canonical solutions the preceding identity gives
\begin{equation}\label{div-g}
\delta\HC\big|_{\mathrm{CT}}=\delta\HC\big|_{\mathrm{fields}}\quad\Leftrightarrow\quad\pfrac{j^{\alpha}}{x^{\alpha}}=0.
\end{equation}
The left-hand equality, required as a phase-space identity, is the
\emph{additional} condition for a canonical transformation to be a symmetry of the given Hamiltonian.
The right-hand equality is then an on-shell conservation law, not an identity for arbitrary field histories.
Conversely, conservation for all local solutions implies the phase-space
identity if every point of the domain lies on such a solution.
Conservation along a single solution does not suffice.
The $j^{\mu}(x)$ then define a conserved four-current vector,
commonly referred to as \emph{Noether current}.
The canonical transformation rules~(\ref{gl1}) then furnish the
corresponding infinitesimal symmetry transformation.
Noether's theorem and its inverse can now be formulated in the
realm of covariant Hamiltonian field theory as:
\newtheorem{ntft}{Theorem}
\begin{ntft}[Hamiltonian Noether]
Let $j^{\mu}(\phi_{I},\Pi_{I},x)$ be admissible in the sense stated above,
and let Eq.~(\ref{gen-infini}) define a locally invertible infinitesimal
canonical transformation. If this transformation preserves the functional
form of a specified Hamiltonian, then its characteristic current has
\emph{zero divergence} on every solution of the canonical field equations.

Conversely, if the admissible characteristic current is conserved on every
local solution, and every point of the relevant phase-space domain lies on
such a solution, then the transformation rules~(\ref{gl1}) define an
infinitesimal symmetry of that Hamiltonian.
\end{ntft}
\section{\label{sec:Noether-u1}Example~1: U$(1)$ gauge theory}
\subsection{Finite symmetry transformation}
As an example, we consider
the covariant Hamiltonian density $\HC_{\mathrm{KGM}}$ of a complex Klein-Gordon $\phi$
field that couples to an electromagnetic $4$-vector potential $a_{\mu}$
\begin{equation}
\HC_{\mathrm{KGM}}=\overline{\pi}_{\alpha}\pi^{\alpha}+m^{2}\overline{\phi}\phi-\quarter p^{\alpha\beta}p_{\alpha\beta}
+\rmi q\left(\overline{\pi}^{\alpha}a_{\alpha}\phi-\overline{\phi}a_{\alpha}\pi^{\alpha}\right),\quad
p^{\alpha\beta}=-p^{\beta\alpha}.
\label{hd-kg-max}
\end{equation}
Herein, the (2,0)-tensor field $p^{\alpha\beta}$ denotes the conjugate
momentum field of $a_{\alpha}$ with a bar denoting complex conjugation.
The complex fields and their conjugates are varied independently, equivalently
to varying their real and imaginary parts.
We assume $q\ne0$ for the gauge transformation formulas.
The dynamics is defined by the first-order action
\begin{displaymath}
S=\int_R\left[
\overline\pi^{\mu}\pfrac{\phi}{x^{\mu}}+
\pfrac{\overline\phi}{x^{\mu}}\pi^{\mu}+
p^{\alpha\beta}\pfrac{a_{\alpha}}{x^{\beta}}
-\HC_{\mathrm{KGM}}\right]\dd^4x,
\end{displaymath}
with antisymmetric variations $\delta p^{\alpha\beta}=-\delta p^{\beta\alpha}$.
The potential and scalar fields have otherwise independent variations.
Only the antisymmetric part of $\partial_{\beta}a_{\alpha}$ couples to $p^{\alpha\beta}$.
Accordingly, transformation rules involving the gauge momenta are read on
this constrained momentum space, not as unrestricted derivatives with
respect to sixteen independent components.
For constraint-preserving symmetries the same Noether argument follows
from the first variation of this action: its boundary current is
$\overline\pi^{\mu}\delta\phi+\delta\overline\phi\,\pi^{\mu}
+p^{\alpha\mu}\delta a_{\alpha}$.
We now define for this Hamiltonian density a \emph{local} symmetry
transformation by means of the generating function
\begin{equation}\label{gen-erwet}
F_{2}^{\mu}=\overline{\Pi}^{\mu}\phi\,e^{\rmi\Lambda(x)}+
\overline{\phi}\,\Pi^{\mu}e^{-\rmi\Lambda(x)}+
P^{\alpha\mu}\left(a_{\alpha}+\frac{1}{q}\pfrac{\Lambda(x)}{x^{\alpha}}\right).
\end{equation}
In this context, the notation ``local'' refers to the fact that
the generating function~(\ref{gen-erwet}) depends \emph{explicitly} on $x$ via $\Lambda=\Lambda(x)$.
The general transformation rules~(\ref{genF2})
applied to the actual generating function, with gauge-momentum
variations restricted as just stated, yield for the fields
\begin{eqnarray}
P^{\mu\nu}&=&p^{\mu\nu},\qquad\qquad\,A_{\mu}=a_{\mu}+\frac{1}{q}\pfrac{\Lambda}{x^{\mu}}\nonumber\\
\Pi^{\mu}&=&\pi^{\mu}e^{\rmi\Lambda(x)},\qquad\;\;\;\Phi=\phi\,e^{\rmi\Lambda(x)}\label{eq:u1-fini}\\
\overline{\Pi}^{\mu}&=&\overline{\pi}^{\mu}\,e^{-\rmi\Lambda(x)},\qquad
\overline{\Phi}=\overline{\phi}\,e^{-\rmi\Lambda(x)}\nonumber
\end{eqnarray}
and for the Hamiltonian from the \emph{explicit} $x^{\mu}$-dependency of $F_{2}^{\mu}$
\begin{eqnarray*}
\HC_{\mathrm{KGM}}^{\prime}-\HC_{\mathrm{KGM}}&=&
\left.\pfrac{F_{2}^{\alpha}}{x^{\alpha}}\right|_{\mathrm{expl}}\\
&=&\rmi\left(\overline{\pi}^{\alpha}\,\phi-
\overline{\phi}\,\pi^{\alpha}\right)\pfrac{\Lambda(x)}{x^{\alpha}}\\
&=&\rmi q\left(\overline{\pi}^{\alpha}\,\phi-
\overline{\phi}\,\pi^{\alpha}\right)\left(A_{\alpha}-a_{\alpha}\right)\\
&=&\rmi q\,\big(\,\overline{\Pi}^{\alpha}A_{\alpha}\Phi-
\overline{\Phi}A_{\alpha}\Pi^{\alpha}\big)-
\rmi q\left(\overline{\pi}^{\alpha}a_{\alpha}\phi-
\overline{\phi}a_{\alpha}\pi^{\alpha}\right).
\end{eqnarray*}
In the transformation rule for the Hamiltonian density, the term
$P^{\alpha\beta}\partial^{2}\Lambda/\partial x^{\alpha}\partial x^{\beta}$
vanishes as the momentum tensor $P^{\alpha\beta}$ is skew-symmetric.
The transformed Hamiltonian density $\HC_{\mathrm{KGM}}^{\prime}$
is now obtained by inserting the transformation rules into the Hamiltonian
density $\HC_{\mathrm{KGM}}$
\begin{displaymath}
\HC_{\mathrm{KGM}}^{\prime}=\overline{\Pi}_{\alpha}\Pi^{\alpha}+m^{2}\overline{\Phi}\Phi-\quarter P^{\alpha\beta}P_{\alpha\beta}
+\rmi q\left(\overline{\Pi}^{\alpha}A_{\alpha}\Phi-\overline{\Phi}A_{\alpha}\Pi^{\alpha}\right).
\end{displaymath}
We observe that the Hamiltonian density~(\ref{hd-kg-max}) is
\emph{form-invariant} under the local canonical transformation
generated by $F_{2}^{\mu}$ from Eq.~(\ref{gen-erwet}) --- which
thus defines a \emph{symmetry transformation} of the given dynamical system.
\subsection{Noether currents and field equations\label{sec:u(1)-eqs}}
In order to derive the conserved Noether current which is
associated with the symmetry transformation~(\ref{eq:u1-fini}),
we first set up the generating function of the \emph{infinitesimal}
canonical transformation corresponding to~(\ref{gen-erwet})
by letting $\Lambda\to\epsilon\Lambda$ and expanding
the exponential function up to the linear term in $\epsilon$.
Equalities in this infinitesimal expansion are understood modulo
$\CO(\epsilon^2)$:
\begin{eqnarray}
F_{2}^{\mu}
&=&\overline{\Pi}^{\mu}\phi(1+\rmi\epsilon\Lambda)+
\overline{\phi}\,\Pi^{\mu}(1-\rmi\epsilon\Lambda)+
P^{\alpha\mu}\left(a_{\alpha}+\frac{\epsilon}{q}\pfrac{\Lambda}{x^{\alpha}}\right)\nonumber\\
&=&\overline{\Pi}^{\mu}\phi+\overline{\phi}\,\Pi^{\mu}+P^{\alpha\mu}a_{\alpha}+
\frac{\epsilon}{q}\left[\rmi q\left(\overline{\pi}^{\mu}\phi-\overline{\phi}\,\pi^{\mu}\right)\Lambda+
p^{\alpha\mu}\pfrac{\Lambda}{x^{\alpha}}\right].
\label{gen-gaugetra}
\end{eqnarray}
According to Noether's theorem~(\ref{div-g}),
the expression in brackets represents the conserved Noether current $j^{\mu}(x)$
\begin{equation}\label{eq:noether-gaugetra}
j^{\mu}(x)=\rmi q\left(\overline{\pi}^{\mu}\phi-
\overline{\phi}\,\pi^{\mu}\right)\Lambda+p^{\beta\mu}\,\pfrac{\Lambda}{x^{\beta}}.
\end{equation}
As the system's symmetry transformation~(\ref{eq:u1-fini}) holds for \emph{arbitrary}
differentiable functions $\Lambda=\Lambda(x)$, the Noether current~(\ref{eq:noether-gaugetra})
must be conserved on solutions for all smooth $\Lambda(x)$.
For the local coefficient argument, $\Lambda$ may be chosen with compact
support, so no boundary restriction prevents its value and derivatives
from being specified independently at an interior point.
The divergence of $j^{\mu}(x)$ is given by:
\begin{eqnarray}
\pfrac{j^{\alpha}}{x^{\alpha}}&=&\Lambda\left[\pfrac{}{x^{\alpha}}\rmi q\left(
\overline{\pi}^{\alpha}\phi-\overline{\phi}\pi^{\alpha}\right)\right]\nonumber\\*
&&\mbox{}+\pfrac{\Lambda}{x^{\beta}}\left[\rmi q\left(\overline{\pi}^{\beta}\phi-
\overline{\phi}\,\pi^{\beta}\right)+\pfrac{p^{\beta\alpha}}{x^{\alpha}}\right]+
\ppfrac{\Lambda}{x^{\beta}}{x^{\alpha}}\,p^{\beta\alpha}.
\label{eq:noether-current}
\end{eqnarray}
With $\Lambda(x)$ an \emph{arbitrary} function of space-time, the divergence
of $j^{\mu}(x)$ vanishes if and only if the three terms associated with
$\Lambda(x)$ and its derivatives in Eq.~(\ref{eq:noether-current}) separately vanish.
This means in particular that the term $j^{\alpha}_{1}$ proportional to $\Lambda$
of the divergence~(\ref{eq:noether-current}) of the Noether current is separately conserved
\begin{equation}\label{eq:conserved-current}
j^{\alpha}_{1}=\rmi q\left(\overline{\pi}^{\alpha}\phi-\overline{\phi}\,\pi^{\alpha}\right),\qquad
\pfrac{j^{\alpha}_{1}}{x^{\alpha}}=0.
\end{equation}
The second term depicts the inhomogeneous Maxwell equation:
\begin{equation}\label{eq:Maxwell}
\pfrac{p^{\beta\alpha}}{x^{\alpha}}=-j^{\beta}_{1}.
\end{equation}
The coefficient of the arbitrary symmetric second derivatives of
$\Lambda$ vanishes precisely when the canonical momentum tensor is
skew-symmetric:
\begin{equation*}
p^{\alpha\beta}=-p^{\beta\alpha}.
\end{equation*}
This checks the antisymmetry already imposed in Eq.~(\ref{hd-kg-max})
and is not a new, independent derivation of that constraint.
Together with Eq.~(\ref{eq:Maxwell}), it gives the consistency requirement:
\begin{displaymath}
\ppfrac{p^{\alpha\beta}}{x^{\alpha}}{x^{\beta}}=\pfrac{j^{\alpha}_{1}}{x^{\alpha}}=0.
\end{displaymath}
The explicit proof of a vanishing divergence of the Noether current $j^{\mu}_{1}$
from Eq.~(\ref{eq:conserved-current}) follows by inserting the canonical
field equations~(\ref{eq:fgln}) emerging from the Hamiltonian~(\ref{hd-kg-max})
\begin{eqnarray*}
\frac{1}{\rmi q}\pfrac{j^{\alpha}_{1}}{x^{\alpha}}&=&\overline{\pi}^{\alpha}\pfrac{\phi}{x^{\alpha}}
-\pfrac{\overline{\phi}}{x^{\alpha}}\pi^{\alpha}+\pfrac{\overline{\pi}^{\alpha}}{x^{\alpha}}\phi
-\overline{\phi}\pfrac{\pi^{\alpha}}{x^{\alpha}}\\
&=&\overline{\pi}^{\alpha}\pfrac{\HC_{\mathrm{KGM}}}{\overline{\pi}^{\alpha}}
-\pfrac{\HC_{\mathrm{KGM}}}{\pi^{\alpha}}\pi^{\alpha}-\pfrac{\HC_{\mathrm{KGM}}}{\phi}\phi
+\overline{\phi}\pfrac{\HC_{\mathrm{KGM}}}{\overline{\phi}}\\
&=&\overline{\pi}^{\alpha}\left(\pi_{\alpha}+\rmi q\,a_{\alpha}\phi\right)
-\left(\overline{\pi}_{\alpha}-\rmi q\,a_{\alpha}\overline{\phi}\right)\pi^{\alpha}\\
&&-\left(\rmi q\,\overline{\pi}^{\alpha}a_{\alpha}+m^{2}\overline{\phi}\right)\phi
+\overline{\phi}\left(m^{2}\phi-\rmi q\,a_{\alpha}\pi^{\alpha}\right)\\
&=&0.
\end{eqnarray*}
Hence, $j^{\alpha}_{1}(x)$ from Eq.~(\ref{eq:conserved-current}) is
conserved along solutions of the canonical field equations for $\HC_{\mathrm{KGM}}$.

On the source equation~(\ref{eq:Maxwell}), the parameter-dependent
current~(\ref{eq:noether-gaugetra}) can also be written as
\begin{displaymath}
j^{\mu}[\Lambda]=
\pfrac{}{x^{\beta}}\left(\Lambda p^{\beta\mu}\right).
\end{displaymath}
It is thus, on shell, the divergence of an antisymmetric superpotential.
For an arbitrary local gauge parameter, the corresponding off-shell
variational relations are Noether identities among the field equations.
This is the setting of Noether's second theorem~\cite{noether18}.
The coefficient conditions above recover the source equation and its
consistency relation on solutions, but do not determine the full dynamics.
In particular, the current does not depend on $a_{\mu}$ and does not fix its momentum--derivative relation.
Antisymmetric variation of $p^{\alpha\beta}$ in the specified first-order action supplies
\begin{displaymath}
\pfrac{a_{\beta}}{x^{\alpha}}-\pfrac{a_{\alpha}}{x^{\beta}}=p_{\alpha\beta}.
\end{displaymath}
The non-Abelian example below exhibits the same distinction between
Noether relations and dynamical input from the Hamiltonian.
\section{\label{sec:Noether-uN}Example~2: U$(N)$ and SU$(N)$ gauge theories}
\subsection{Finite symmetry transformation}
Similarly to the U$(1)$ case of Eq.~(\ref{hd-kg-max}), we consider
$N$ complex scalar fields and the Yang-Mills Hamiltonian $\HC_{\mathrm{YM}}$.
We first work with U$(N)$, the additional restrictions needed for SU$(N)$ are stated below.
The matrices $a_{\mu}=(a_{JK\mu})$ and
$p^{\alpha\beta}=(p_{JK}^{\alpha\beta})$ are Hermitian, with
\mbox{$p_{JK}^{\alpha\beta}=-p_{JK}^{\beta\alpha}$}.
The Hamiltonian
\begin{eqnarray*}
\HC_{\mathrm{YM}}&=&\overline{\pi}_{J\alpha}\pi_{J}^{\alpha}+m^{2}\,
\overline{\phi}_{J}\phi_{J}-\quarter p_{JK}^{\alpha\beta}\,p_{KJ\alpha\beta}\\
&&\mbox{}+\rmi q\left(\overline{\pi}_{K}^{\alpha}\,a_{KJ\alpha}\,\phi_{J}-
\overline{\phi}_{K}\,a_{KJ\alpha}\,\pi_{J}^{\alpha}-
p_{JK}^{\alpha\beta}\,a_{KI\alpha}\,a_{IJ\beta}\right)
\end{eqnarray*}
can be shown to be form-invariant under the \emph{local} transformations below.
Its matter Hamiltonian,
\begin{displaymath}
\HC\big(\phi_{I},\overline{\phi}_{I},\pi_{I}^{\mu},\overline{\pi}_{I}^{\mu}\big)=
\overline{\pi}_{J\alpha}\pi_{J}^{\alpha}+m^{2}\,\overline{\phi}_{J}\phi_{J},
\end{displaymath}
is invariant under the corresponding \emph{global} transformation
\begin{displaymath}
\Phi_{I}=u_{IJ}\,\phi_{J},\quad\overline{\Phi}_{I}=\overline{\phi}_{J}\,u_{JI}^{*}.
\end{displaymath}
The $u_{IJ}$ are the entries of a \emph{unitary} matrix $U$.
To retain the component notation, we define
$u_{IJ}^{*}:=(U^{\dagger})_{IJ}=\overline{u_{JI}}$.
The star on a matrix entry thus denotes the corresponding entry of the
Hermitian adjoint, not entrywise conjugation at fixed indices.
Unitarity then gives
\begin{displaymath}
u_{JI}^{*}\,u_{IK}=\delta_{JK}=u_{JI}\,u_{IK}^{*}.
\end{displaymath}
At this point, the unitary matrix $U=(u_{IJ})$ is usually expressed in textbooks in terms of its representation
\begin{equation}\label{eq:U-rep}
U=\exp\left(\frac{\rmi}{2}\btau\cdot\balpha\right),
\end{equation}
where $\balpha$ denotes the real vector of group parameters --- the
counterpart of the phase $\Lambda$ of U$(1)$ gauge theory.
There are $N^2$ components for U$(N)$ and $N^2-1$ for SU$(N)$.
The $N\times N$ Hermitian matrices $\btau$ form a generator basis.
For SU$(N)$ they are also traceless (Pauli matrices, Gell-Mann matrices,
\ldots), whereas U$(N)$ includes an identity generator.
Yet, for the sake of simplicity of the derivation, we do not pursue this formulation here,
but continue to work with the coefficients $u_{IJ}$.
Their particular representation~(\ref{eq:U-rep}) can be inserted at any point later in the derivation.
On the other hand, it is the spirit of all gauge theories to finally replace all dependencies on the
arbitrary coefficients of a particular symmetry transformation by gauge fields,
which finally yields a Lagrangian/Hamiltonian completely independent of those coefficients.
For this reason, there is no need to specify an explicit representation of the unitary matrix $U=(u_{IJ})$ in the actual context.

As in the Abelian example, the dynamics is defined by the first-order action
\begin{displaymath}
S=\int_R\left[
\overline\pi_J^{\mu}\pfrac{\phi_J}{x^{\mu}}+
\pfrac{\overline\phi_J}{x^{\mu}}\pi_J^{\mu}+
p_{JK}^{\alpha\beta}\pfrac{a_{KJ\alpha}}{x^{\beta}}
-\HC_{\mathrm{YM}}\right]\dd^4x.
\end{displaymath}
Gauge-field variations preserve Hermiticity, and gauge-momentum variations
also preserve antisymmetry in the space-time indices.
The generating function of the \emph{local} symmetry transformation is given by
\begin{equation}\label{eq:gen-gaugetra}
F_{2}^{\mu}=\overline{\Pi}_{K}^{\mu}\,u_{KJ}\,\phi_{J}+
\overline{\phi}_{K}\,u^{*}_{KJ}\,\Pi_{J}^{\mu}+P_{JK}^{\alpha\mu}
\left(u_{KL}\,a_{LI\alpha}\,u^{*}_{IJ}+\frac{1}{\rmi q}
\pfrac{u_{KI}}{x^{\alpha}}\,u^{*}_{IJ}\right).
\end{equation}
It entails the canonical transformation rules for the complex fields and their conjugates
\begin{eqnarray}
\overline{\pi}_{I}^{\mu}&=&\overline{\Pi}_{K}^{\mu}\,u_{KI},\qquad
\overline{\Phi}_{I}=\overline{\phi}_{K}u^{*}_{KI}\nonumber\\
\pi_{I}^{\mu}&=&u^{*}_{IJ}\Pi_{J}^{\mu},\qquad\;\;\Phi_{I}=u_{IJ}\phi_{J}
\label{eq:symm1}
\end{eqnarray}
and the following rules for the Hermitian $N\times N$ matrix of $4$-vector gauge fields $a_{LI\alpha}$ and their conjugates
\begin{eqnarray}
A_{KJ\alpha}&=&u_{KL}\,a_{LI\alpha}\,u^{*}_{IJ}+\frac{1}{\rmi q}
\pfrac{u_{KI}}{x^{\alpha}}u^{*}_{IJ}\nonumber\\
p_{IL}^{\alpha\mu}&=&u^{*}_{IJ}P_{JK}^{\alpha\mu}\,u_{KL}.
\label{eq:symm2}
\end{eqnarray}
The explicit dependence of the generating function~(\ref{eq:gen-gaugetra})
on $x$ yields the Hamiltonian transformation
\begin{displaymath}
\HC_{\mathrm{YM}}^{\prime}-\HC_{\mathrm{YM}}=\left.\pfrac{F_{2}^{\alpha}}{x^{\alpha}}\right|_{\mathrm{expl}}.
\end{displaymath}
Expressing all $u_{IJ}$-dependent terms in this equation in terms of the fields
and their conjugates according to the above canonical transformation
rules~(\ref{eq:symm1}) and~(\ref{eq:symm2}) finally yields~\cite{StrRei12}
\begin{eqnarray*}
\HC_{\mathrm{YM}}^{\prime}-\HC_{\mathrm{YM}}&=&
\rmi q\,\Big[\overline{\Pi}_{K}^{\alpha}A_{KJ\alpha}\Phi_{J}-
\overline{\Phi}_{K}A_{KJ\alpha}\Pi_{J}^{\alpha}-P_{JK}^{\alpha\beta}A_{KI\alpha}\,A_{IJ\beta}\nonumber\\
&&\quad\,\,\mbox{}-\left(\overline{\pi}_{K}^{\alpha}a_{KJ\alpha}\phi_{J}-
\overline{\phi}_{K}a_{KJ\alpha}\pi_{J}^{\alpha}-p_{JK}^{\alpha\beta}a_{KI\alpha}\,a_{IJ\beta}\right)\Big].
\end{eqnarray*}
Again, we made use of the fact that the momentum fields $p_{JK}^{\alpha\beta}$
are skew-symmetric in $\alpha,\beta$.
The transformed Hamiltonian now follows with $P_{JK}^{\mu\nu}=-P_{JK}^{\nu\mu}$ as
\begin{eqnarray*}
\HC_{\mathrm{YM}}^{\prime}&=&\overline{\Pi}_{J\alpha}\Pi_{J}^{\alpha}+m^{2}\,
\overline{\Phi}_{J}\Phi_{J}-\quarter P_{JK}^{\alpha\beta}\,P_{KJ\alpha\beta}\\
&&\mbox{}+\rmi q\left(\overline{\Pi}_{K}^{\alpha}\,A_{KJ\alpha}\,\Phi_{J}-
\overline{\Phi}_{K}\,A_{KJ\alpha}\,\Pi_{J}^{\alpha}-
P_{JK}^{\alpha\beta}\,A_{KI\alpha}\,A_{IJ\beta}\right),
\end{eqnarray*}
which has the same form as the original one, $\HC_{\mathrm{YM}}$.
Thus, the generating function~(\ref{eq:gen-gaugetra}) defines
a \emph{local symmetry transformation} of the Yang-Mills Hamiltonian.
The transformations preserve the stated momentum constraints.
Substitution in the first-order action gives the same canonical action
identity on the constrained phase space.
\subsection{Noether currents and field equations}
In order to derive the conserved Noether current which is associated with the
symmetry transformation given by Eqs.~(\ref{eq:symm1}) and~(\ref{eq:symm2}),
we again set up the generating function of the corresponding \emph{infinitesimal}
transformation by writing
\begin{displaymath}
U=\mathbb{1}_N+\rmi\epsilon\,\omega+\CO(\epsilon^2),\qquad
U^{\dagger}=\mathbb{1}_N-\rmi\epsilon\,\omega+\CO(\epsilon^2),
\end{displaymath}
hence, to first order,
\begin{displaymath}
\Phi_{I}=\left(\delta_{IJ}+\rmi\epsilon\,\omega_{IJ}\right)\phi_{J},\qquad
\overline{\Phi}_{I}=\overline{\phi}_{J}\left(\delta_{JI}-\rmi\epsilon\,\omega_{JI}\right).
\end{displaymath}
For the \emph{local} transformation, $\omega=\omega(x)$ is an arbitrary smooth Hermitian matrix.
Its entries need not be real; its coefficients in a Hermitian generator basis are real.
For SU$(N)$ one additionally requires $\operatorname{tr}\omega=0$, since
$\det U=1+\rmi\epsilon\operatorname{tr}\omega+\CO(\epsilon^2)$.
There is no condition $\det\omega=1$.
The generating function~(\ref{eq:gen-gaugetra}) is then expanded into
the generating function of the corresponding \emph{infinitesimal} canonical transformation
\begin{eqnarray*}
F_{2}^{\mu}&=&\overline{\Pi}_{K}^{\mu}\left(\delta_{KJ}+\rmi\epsilon\,\omega_{KJ}\right)\phi_{J}+
\overline{\phi}_{K}\left(\delta_{KJ}-\rmi\epsilon\,\omega_{KJ}\right)\Pi_{J}^{\mu}\\
&&\quad\mbox{}+P_{JK}^{\alpha\mu}\bigg[\left(\delta_{KL}+\rmi\epsilon\,\omega_{KL}\right)a_{LI\alpha}
\left(\delta_{IJ}-\rmi\epsilon\,\omega_{IJ}\right)+\frac{\epsilon}{q}\pfrac{\omega_{KI}}{x^{\alpha}}
\left(\delta_{IJ}-\rmi\epsilon\,\omega_{IJ}\right)\bigg].
\end{eqnarray*}
Omitting the quadratic terms in $\epsilon$, the generating function of the
sought-for infinitesimal canonical transformation is obtained as
\begin{equation}\label{gen-gaugetra-infini}
F_{2}^{\mu}=\overline{\Pi}_{J}^{\mu}\phi_{J}+\overline{\phi}_{J}\Pi_{J}^{\mu}+
P_{JK}^{\alpha\mu}\,a_{KJ\alpha}+\frac{\epsilon}{q}j^{\mu},
\end{equation}
with the Noether current of the U$(N)$ gauge theory
\begin{equation}
j^{\mu}=\rmi q\left[\overline{\pi}_{K}^{\mu}\,\omega_{KJ}\phi_{J}-
\overline{\phi}_{K}\,\omega_{KJ}\pi_{J}^{\mu}\vphantom{\pfrac{\omega_{KJ}}{x^{\alpha}}}+p_{JK}^{\alpha\mu}\left(
\omega_{KI}a_{IJ\alpha}-a_{KI\alpha}\omega_{IJ}+\frac{1}{\rmi q}\pfrac{\omega_{KJ}}{x^{\alpha}}\right)\right].
\label{noether-gen-gaugetra}
\end{equation}
Since this generator describes an infinitesimal Hamiltonian symmetry,
Noether's theorem gives $\partial j^{\beta}/\partial x^{\beta}=0$ on
solutions for every smooth Hermitian $\omega(x)$.
Calculating its divergence and ordering the terms according to zeroth,
first and second derivatives of the $\omega_{KJ}(x)$ yields
\begin{eqnarray}
\frac{1}{\rmi q}\pfrac{j^{\beta}}{x^{\beta}}&=&\omega_{KJ}\pfrac{}{x^{\beta}}\left(
\overline{\pi}_{K}^{\beta}\phi_{J}-\overline{\phi}_{K}\pi_{J}^{\beta}+
a_{JI\alpha}p_{IK}^{\alpha\beta}-p_{JI}^{\alpha\beta}a_{IK\alpha}\right)\nonumber\\
&&\mbox{}+\pfrac{\omega_{KJ}}{x^{\beta}}\left(\overline{\pi}_{K}^{\beta}\phi_{J}-\overline{\phi}_{K}\pi_{J}^{\beta}+
a_{JI\alpha}p_{IK}^{\alpha\beta}-p_{JI}^{\alpha\beta}a_{IK\alpha}+
\frac{1}{\rmi q}\pfrac{p_{JK}^{\beta\alpha}}{x^{\alpha}}\right)\nonumber\\
&&\mbox{}+\frac{1}{\rmi q}\ppfrac{\omega_{KJ}}{x^{\alpha}}{x^{\beta}}p_{JK}^{\alpha\beta}.
\label{eq:noether-current-gen}
\end{eqnarray}
The $N^2$ real coefficient functions of $\omega(x)$ in a Hermitian
U$(N)$ generator basis may be chosen independently, together with their
first and symmetric second derivatives at a point.
The corresponding coefficients in the divergence therefore vanish separately on solutions.
The second-derivative term vanishes by the already imposed antisymmetry of $p^{\alpha\beta}$.
This means in particular that the term $j^{\mu}_{JK}$ proportional to $\omega_{KJ}$
of the divergence of the Noether current~(\ref{eq:noether-current-gen}) is separately conserved
\begin{equation}\label{eq:color-current-expl}
j^{\mu}_{JK}=\rmi q\left(\phi_{J}\,\overline{\pi}_{K}^{\mu}-\pi_{J}^{\mu}\,\overline{\phi}_{K}+
a_{JI\alpha}\,p_{IK}^{\alpha\mu}-p_{JI}^{\alpha\mu}\,a_{IK\alpha}\right),\qquad
\pfrac{j^{\beta}_{JK}}{x^{\beta}}=0,
\end{equation}
whereas the first-derivative coefficient, together with the imposed antisymmetry,
\begin{equation}\label{eq:color-current}
\pfrac{p^{\alpha\mu}_{JK}}{x^{\alpha}}=j^{\mu}_{JK},\qquad p^{\alpha\mu}_{JK}=-p^{\mu\alpha}_{JK},
\end{equation}
is the U$(N)$ generalization of the Maxwell equation~(\ref{eq:Maxwell})
which similarly satisfies the consistency requirement
\begin{displaymath}
\ppfrac{p^{\alpha\beta}_{JK}}{x^{\alpha}}{x^{\beta}}=-\ppfrac{p^{\beta\alpha}_{JK}}{x^{\alpha}}{x^{\beta}}=
\pfrac{j^{\beta}_{JK}}{x^{\beta}}=0.
\end{displaymath}
The $j^{\mu}_{JK}$ define the ordinarily conserved \emph{total} U$(N)$ currents in this representation.
They include the gauge-field contribution and are not themselves gauge-covariant matter currents.
The source equation~(\ref{eq:color-current}) also follows directly by
varying $a_{KJ\mu}$ in the specified first-order action.
On this equation, the parameter-dependent Noether current becomes
\begin{displaymath}
j^{\mu}[\omega]=\pfrac{}{x^{\alpha}}
\left(\omega_{KJ}p_{JK}^{\alpha\mu}\right),
\end{displaymath}
an antisymmetric-superpotential divergence, as in the Abelian example.
In contrast to the Abelian case, the fields $a_{JK}^{\mu}$ themselves contribute to the source terms $j_{JK}^{\mu}$,
which is referred to as the ``self-coupling effect'' of non-Abelian gauge theories.
The explicit representation of the divergence of the U$(N)$ gauge currents~(\ref{eq:color-current-expl}) evaluates to
\begin{eqnarray*}
\frac{1}{\rmi q}\pfrac{j^{\beta}_{JK}}{x^{\beta}}&=&\pfrac{}{x^{\beta}}\left(\phi_{J}\overline{\pi}_{K}^{\beta}-
\pi_{J}^{\beta}\overline{\phi}_{K}+a_{JI\alpha}p_{IK}^{\alpha\beta}-p_{JI}^{\alpha\beta}a_{IK\alpha}\right)\\
&=&\pfrac{}{x^{\beta}}\left(\phi_{J}\overline{\pi}_{K}^{\beta}-\pi_{J}^{\beta}\overline{\phi}_{K}\right)+
\pfrac{a_{JI\alpha}}{x^{\beta}}p_{IK}^{\alpha\beta}\\
&&\mbox{}-a_{JI\alpha}j^{\alpha}_{IK}+j^{\alpha}_{JI}a_{IK\alpha}-
p_{JI}^{\alpha\beta}\pfrac{a_{IK\alpha}}{x^{\beta}},
\end{eqnarray*}
where the divergences of the momenta $p_{IK}^{\alpha\beta}$ were replaced
by the U$(N)$ gauge currents $j_{IK}^{\alpha}$ according to Eq.~(\ref{eq:color-current}).
Inserting finally the explicit representation~(\ref{eq:color-current-expl}) of the U$(N)$ gauge currents yields
\begin{eqnarray}
0&=&\frac{1}{\rmi q}\pfrac{j^{\beta}_{JK}}{x^{\beta}}\nonumber\\
&=&\overline{\pi}_{K}^{\alpha}\left(\pfrac{\phi_{J}}{x^{\alpha}}-\rmi q\,a_{JI\alpha}\phi_{I}\right)-
\left(\pfrac{\overline{\phi}_{K}}{x^{\alpha}}+\rmi q\,\overline{\phi}_{I}a_{IK\alpha}\right)\pi_{J}^{\alpha}\nonumber\\
&&\mbox{}+\left(\pfrac{\overline{\pi}_{K}^{\alpha}}{x^{\alpha}}+\rmi q\,\overline{\pi}_{I}^{\alpha}a_{IK\alpha}\right)\phi_{J}-
\overline{\phi}_{K}\left(\pfrac{\pi_{J}^{\alpha}}{x^{\alpha}}-\rmi q\,a_{JI\alpha}\pi_{I}^{\alpha}\right)\nonumber\\
&&\mbox{}+\onehalf\left[\pfrac{a_{JI\alpha}}{x^{\beta}}-\pfrac{a_{JI\beta}}{x^{\alpha}}+\rmi q\left(
a_{JN\alpha}a_{NI\beta}-a_{JN\beta}a_{NI\alpha}\right)\right]p_{IK}^{\alpha\beta}\nonumber\\
&&\mbox{}-\onehalf p_{JI}^{\alpha\beta}\left[\pfrac{a_{IK\alpha}}{x^{\beta}}-\pfrac{a_{IK\beta}}{x^{\alpha}}+
\rmi q\left(a_{IN\alpha}a_{NK\beta}-a_{IN\beta}a_{NK\alpha}\right)\right]\label{eq:qcd-noether}
\end{eqnarray}
Equation~(\ref{eq:qcd-noether}) is a consistency relation.
It does not imply the complete canonical equations.
To make the additional dynamical input explicit, recall the free,
\emph{globally} form-invariant Klein-Gordon Hamiltonian used in the chosen model,
\begin{displaymath}
\HC=\overline{\pi}_{J\alpha}\pi_{J}^{\alpha}+m^{2}\, \overline{\phi}_{J}\phi_{J},
\end{displaymath}
hence
\begin{eqnarray*}
\pfrac{\phi_{J}}{x^{\alpha}}&=&\pfrac{\HC}{\overline{\pi}_{J}^{\alpha}}=\pi_{J\alpha},\qquad
\pfrac{\pi_{J}^{\alpha}}{x^{\alpha}}=-\pfrac{\HC}{\overline{\phi}_{J}}=-m^{2}\phi_{J}\\
\pfrac{\overline{\phi}_{J}}{x^{\alpha}}&=&\pfrac{\HC}{\pi_{J}^{\alpha}}=\overline{\pi}_{J\alpha},\qquad
\pfrac{\overline{\pi}_{J}^{\alpha}}{x^{\alpha}}=-\pfrac{\HC}{\phi_{J}}=-m^{2}\overline{\phi}_{J}.
\end{eqnarray*}
Sufficient conditions for Eq.~(\ref{eq:qcd-noether}) to vanish are the
\emph{amended} canonical equations of the \emph{locally} form-invariant system,
obtained by varying the scalar fields and their momenta in the first-order
action with the displayed $\HC_{\mathrm{YM}}$,
\begin{eqnarray}
\pi_{J\alpha}&=&\pfrac{\phi_{J}}{x^{\alpha}}-\rmi q\,a_{JI\alpha}\phi_{I}\nonumber\\
\overline{\pi}_{K\alpha}&=&\pfrac{\overline{\phi}_{K}}{x^{\alpha}}+\rmi q\,\overline{\phi}_{I}a_{IK\alpha}\nonumber\\
\pfrac{\pi_{J}^{\alpha}}{x^{\alpha}}&=&-m^{2}\phi_{J}+\rmi q\,a_{JI\alpha}\pi_{I}^{\alpha}\nonumber\\
\pfrac{\overline{\pi}_{K}^{\alpha}}{x^{\alpha}}&=&-m^{2}\overline{\phi}_{K}-\rmi q\,\overline{\pi}_{I}^{\alpha}a_{IK\alpha}\label{eq:pi-phi-corr}
\end{eqnarray}
together with the momentum--curvature relation.
In matrix notation, antisymmetric variation of $p^{\alpha\beta}$ gives
\begin{displaymath}
\pfrac{a_{\alpha}}{x^{\beta}}-\pfrac{a_{\beta}}{x^{\alpha}}
=-p_{\alpha\beta}-\rmi q\left(a_{\alpha}a_{\beta}-a_{\beta}a_{\alpha}\right).
\end{displaymath}
Equivalently, in the original component notation,
\begin{equation}\label{eq:p-a-corr}
p_{JI\beta\alpha}=\pfrac{a_{JI\alpha}}{x^{\beta}}-\pfrac{a_{JI\beta}}{x^{\alpha}}+\rmi q\left(
a_{JN\alpha}a_{NI\beta}-a_{JN\beta}a_{NI\alpha}\right).
\end{equation}
The canonical momenta $\pi_{J\alpha}$ and $\overline{\pi}_{K\alpha}$ turn out to represent
the gauge-covariant derivatives of the pertaining fields $\phi_{J}$ and $\overline{\phi}_{K}$, respectively.
In conjunction with Eqs.~(\ref{eq:color-current-expl}) and~(\ref{eq:color-current}),
these equations specify the dynamics of the chosen Hamiltonian system,
with the usual gauge freedom and subject to appropriate initial and boundary data.
Substitution into Eq.~(\ref{eq:qcd-noether}) makes its
right-hand side vanish, but that implication cannot in general be reversed.
In particular, the quadratic momentum term of $\HC_{\mathrm{YM}}$ is
additional input: gauge invariance and a prescribed free-matter limit do
not uniquely determine the gauge-field kinetic Hamiltonian or the
interacting matter potential~\cite{StrRei12,koenigstein16}.

\paragraph{Restriction to SU$(N)$.}
For SU$(N)$, $N\geq2$, the finite transformations must satisfy $\det U=1$,
and the gauge variables and their variations are also traceless:
\begin{displaymath}
\operatorname{tr}a_{\mu}=0,\qquad
\operatorname{tr}p^{\alpha\beta}=0,\qquad
\operatorname{tr}\omega=0.
\end{displaymath}
The generating function and field transformations above restrict consistently to this space.
Since an arbitrary traceless Hermitian parameter probes only the traceless part
of a matrix coefficient, the source equation and its consistency relation become
\begin{displaymath}
(j_{\mathrm{SU}}^{\mu})_{JK}=j_{JK}^{\mu}-\frac{\delta_{JK}}{N}
j_{LL}^{\mu},\qquad
\pfrac{p^{\alpha\mu}}{x^{\alpha}}=j_{\mathrm{SU}}^{\mu},\qquad
\pfrac{j_{\mathrm{SU}}^{\mu}}{x^{\mu}}=0,
\end{displaymath}
where $j_{JK}^{\mu}$ is the total-current expression of
Eq.~(\ref{eq:color-current-expl}) and $j_{\mathrm{SU}}^{\mu}$ denotes the
matrix with the displayed projected components.
The trace of the commutator contribution vanishes, so it is the matter source that must be projected.
Equivalently,
\begin{displaymath}
\mathcal{D}_{\alpha}p^{\alpha\mu}
=j_{\mathrm{mat}}^{\mu}-\frac{1}{N}
\operatorname{tr}(j_{\mathrm{mat}}^{\mu})\mathbb{1}_N,\qquad
\mathcal{D}_{\alpha}X=\partial_{\alpha}X-\rmi q[a_{\alpha},X],
\end{displaymath}
with $(j_{\mathrm{mat}}^{\mu})_{JK}
=\rmi q(\phi_J\overline\pi_K^{\mu}-\pi_J^{\mu}\overline\phi_K)$.
Unlike the total current, this projected matter source is gauge-covariant.
Equation~(\ref{eq:qcd-noether}) is correspondingly read after traceless projection.
Eqs.~(\ref{eq:pi-phi-corr}) and~(\ref{eq:p-a-corr}) otherwise retain their form.
In particular, no condition that the unprojected
matter current have zero trace is imposed.

\paragraph{Abelian reduction.}
The momentum--curvature relation~(\ref{eq:p-a-corr}), obtained from the
chosen kinetic Hamiltonian, also includes the Abelian result.
Restricting the range of the field indices to $I=J=N=1$---hence to \emph{one} (real) gauge field
$a_{\mu}\equiv a_{11\mu}$ and thus \emph{one} canonical momentum tensor $p^{\mu\nu}\equiv p_{11}^{\mu\nu}$
--- corresponds to the reduction U$(N)\to$U$(1)$.
As only the self-coupling terms cancel for this case, we get
\begin{equation}\label{eq:p-a-correlation}
p_{\beta\alpha}=\pfrac{a_{\alpha}}{x^{\beta}}-\pfrac{a_{\beta}}{x^{\alpha}},
\end{equation}
which agrees with the result of the antisymmetric variation in Sect.~\ref{sec:u(1)-eqs}.
This is the reduction U$(N)\to$U$(1)$ with SU$(1)$ the trivial group.
As a consequence of Eq.~(\ref{eq:p-a-correlation}), one encounters the \emph{homogeneous} Maxwell equation:
\begin{displaymath}
\pfrac{p_{\nu\mu}}{x^{\alpha}}+\pfrac{p_{\mu\alpha}}{x^{\nu}}+\pfrac{p_{\alpha\nu}}{x^{\mu}}=0.
\end{displaymath}
Thus the homogeneous Maxwell equation is the differential identity
associated with the action-derived momentum--curvature relation, not an
additional dynamical consequence of current conservation alone.
\section{Conclusions and outlook}
Our presentation shows how admissible infinitesimal canonical symmetries
of a specified Hamiltonian generate conserved Noether currents.
For the local U$(1)$ and U$(N)$ transformations, the arbitrary gauge
parameters organize the source equations and their consistency relations.
SU$(N)$ requires the corresponding traceless projection.
The full field equations, including the relation of the gauge momenta to
the curvature, follow from the specified first-order action with
antisymmetric gauge-momentum variations.
They are not determined by current conservation or local gauge symmetry alone.
In particular, the gauge-field kinetic Hamiltonian remains dynamical input.
Koenigstein et al.~\cite{koenigstein16} discuss an alternative
canonical approach to U$(1)$ gauge theory and the role of this input.

The covariant canonical approach also admits a generalization in which space-time geometry is dynamical.
Requiring form-invariance under local space-time transformations constrains the couplings and gives the
associated Noether identities.
The dynamics of the geometry additionally requires a gravitational Hamiltonian~\cite{struckvasak16}.
A tetrad formulation includes fermions and their coupling to curvature and torsion~\cite{vasakstruck16}.
Here too, symmetry constraints and the choice of the free gravitational dynamics must be distinguished.
\begin{acknowledgement}
This paper is prepared for the \emph{Symposium on Exciting Physics},
which was held in November~2015 at Makutsi Safari Farm, South Africa,
to honor our teacher, mentor, and friend Prof.~Dr.~Dr.~h.c.~mult.\
Walter Greiner on the occasion of his 80th birthday.
We thank Walter for stimulating generations of young scientists for more than 100~semesters,
both at the Goethe Universit\"{a}t Frankfurt am Main and internationally.
We wish him good health to further take part in the progress of physics in the years to come.

We furthermore thank the present members of our FIAS working group on the \emph{Extended canonical formalism of field theory},
namely Michail Chabanov, Matthias Hanauske, Johannes Kirsch, Adrian Koenigstein, and
Johannes Muench for many fruitful discussions.
\end{acknowledgement}
%\bibliography{/u/struck/doc/book/extLagHam}

\begin{thebibliography}{10}
\providecommand{\url}[1]{{#1}}
\providecommand{\urlprefix}{URL }
\expandafter\ifx\csname urlstyle\endcsname\relax
  \providecommand{\doi}[1]{DOI \discretionary{}{}{}#1}\else
  \providecommand{\doi}{DOI \discretionary{}{}{}\begingroup
  \urlstyle{rm}\Url}\fi

\bibitem{StrRei12}
J.~Struckmeier, H.~Reichau, \emph{General U$(N)$ gauge transformations in the
  realm of covariant {H}amiltonian field theory, in: {E}xciting
  {I}nterdisciplinary {P}hysics}.
\newblock {FIAS} {I}nterdisciplinary {S}cience {S}eries (Springer, New York,
  2013).
\newblock \urlprefix\url{https://arxiv.org/abs/1205.5754}.
\newblock P. 367

\bibitem{noether18}
E.~Noether, \emph{Invariante Variationsprobleme},
  Nachr. K\"{o}nigl. Ges. Wiss. G\"{o}ttingen, Math.-Phys. Kl.
  \textbf{1918}, 235--257 (1918)

\bibitem{saletan98}
J.V. Jos\'e, E.J. Saletan, \emph{Classical Dynamics} (Cambridge University
  Press, Cambridge, 1998)

\bibitem{struckmeier08}
J.~Struckmeier, A.~Redelbach, \emph{Covariant Hamiltonian Field Theory}, Int. J. Mod. Phys. E \textbf{17}, 435 (2008).
\newblock \urlprefix\url{https://arxiv.org/abs/0811.0508}

\bibitem{greiner98}
W.~Greiner, \emph{Classical Electrodynamics} (Springer, 1998)

\bibitem{dedonder30}
T.~De~Donder, \emph{Th\'eorie Invariantive Du Calcul des Variations}
  (Gauthier-Villars \& Cie., Paris, 1930)

\bibitem{weyl35}
H.~Weyl, Annals of Mathematics \textbf{36}, 607 (1935)

\bibitem{koenigstein16}
A.~Koenigstein, J.~Kirsch, H.~Stoecker, J.~Struckmeier, D.~Vasak, M.~Hanauske,
  Int. J. Mod. Phys. E \textbf{25}, 1642005 (2016).
\newblock \doi{10.1142/S0218301316420052}

\bibitem{struckvasak16}
J.~Struckmeier, J.~Muench, D.~Vasak, J.~Kirsch, M.~Hanauske, H.~Stoecker,
  \emph{Canonical transformation path to gauge theories of gravity},
  Phys. Rev. D \textbf{95}, 124048 (2017).
\newblock \doi{10.1103/PhysRevD.95.124048}

\bibitem{vasakstruck16}
J.~Struckmeier, D.~Vasak,
  \emph{Covariant canonical gauge theory of gravitation for fermions},
  Astron. Nachr. \textbf{342}, 745--764 (2021).
\newblock \doi{10.1002/asna.202113991}

\bibitem{vignolo05}
S.~Vignolo, R.~Cianci, D.~Bruno,
  \emph{On the Hamiltonian formulation of Yang--Mills gauge theories},
  Int. J. Geom. Methods Mod. Phys. \textbf{2}, 1115 (2005).
\newblock \urlprefix\url{https://arxiv.org/abs/math-ph/0507001}

\end{thebibliography}
%\input{makutsi15.bbl}

\end{document}